# Personalized Highway Pilot Assist Considering Leading Vehicle's Lateral Behaviours

Daofei Li*[1] and Ao Liu[1]

**Abstract**: Highway pilot assist has become the front line of competition in advanced driver assistance systems. The increasing requirements on safety and user acceptance are calling for personalization in the development process of such systems. Inspired by a finding on drivers' car-following preferences on lateral direction, a personalized highway pilot assist algorithm is proposed, which consists of an Intelligent Driver Model (IDM) based speed control model and a novel lane-keeping model considering the leading vehicle's lateral movement. A simulated driving experiment is conducted to analyse drivers' gaze and lane-keeping behaviours in free-driving and following driving scenario. Drivers are clustered into two driving style groups referring to their driving behaviours affected by the leading vehicle, and then the personalization parameters for every specific subject driver are optimized. The proposed algorithm is validated through driver-in-the-loop experiment based on a moving-base simulator. Results show that, compared with the un-personalized algorithms, the personalized highway pilot algorithm can significantly reduce the mental workload and improve user acceptance of the assist functions.

## 1. Introduction

Recently, automated driving has attracted the attention of both academia and industries around the world. Tech giants, such as Google and Baidu, have launched their SAE Level 4 automated-driving cars for RoboTaxi services. Advanced Driving Assistance Systems (ADAS) have also made great progresses in mass-produced passenger and commercial vehicles. By contributing to safety and comfort, such automated driving functions of SAE Levels 1 or 2, e.g., adaptive cruise control and lane keeping assist, almost become standard features in passenger vehicle models above B-class. Among the high-end ADAS, automated (or even autonomous) highway driving is the front line of the competition. In the near future, Highway Pilot or Hands-free highway driving, as an intelligent combination of both longitudinal and lateral driving functions on highway, is hopefully to prevail in premium brands. As the name suggests, Highway Pilot automates such functions as car following, lane keeping, lane exiting and entering, etc. However, there are still only limited open reports on how such high-end automated driving function is developed.

### 1.1. Longitudinal and lateral automation

For longitudinal driving automation, car following models have been studied and successfully applied in ADAS developments, e.g., adaptive cruise control. The concept of "car-following" can be traced back to the 1950s, a certain prescribed "following distance" was considered in the dynamics of vehicle platoon, i.e. the regulation of safe distance suggested that the following distance needs to be increased by a length of a car for every 10 miles per hour increase in speed [1]. General Motors (GM) has made a great contribution in this field, e.g. the original GM model describes the relationship between the relative speed and the acceleration of the following car [2]. With many improvements by the research community, the GM model has become the most important and classic car following model. Different from the perspective of kinematics and traffic engineering, further studies start to consider drivers' car-following behaviours, e.g., perception and decision-making. Michaels [3] believes that the driver's car following behaviour is based on the size of the leading car in the field of view, i.e. once the size change exceeds a threshold, the driver responds to accelerate or decelerate. This psychological view is also proved by experiments [4]. Since the 1990s, car following theory has entered a new field with statistical modelling, with examples as Optimal Velocity model (OV) [5], Generalized Force model (GF) [6] and Intelligent Driver Model (IDM) [7]. These theories try to describe the nonlinear characteristics of macro traffic flow by modelling micro car following behaviour, and eventually to explain the formation and dissipation mechanism of traffic congestion. Recently, several machine learning based approaches are proposed to model the car following behaviour [8–10]. For example, with feature extraction and self-learning from data, neural network has been extensively applied to car following behaviour modelling.

[1] Institute of Power Machinery and Vehicular Engineering, Faculty of Engineering, Zhejiang University.
* Corresponding author: Daofei Li, Institute of Power Machinery and Vehicular Engineering, Zhejiang University, No 38 Zheda Road, Xihu District, Hangzhou, 310028, China. Email: dfli@zju.edu.cn

Understandably, the car-following models, as the name implies, should basically describe the driver's longitudinal control behaviours when following another car, e.g., how to choose the expected gap, speed or acceleration, etc. In most car-following models, the lateral offset between the leading and following vehicles is not well considered. To this end, in a car following model proposed by Song et al, the longitudinal acceleration of the following vehicle includes the influence of two-vehicle lateral offset[11]. Another car following model has been developed, in which the following vehicle is formulated as a function of the off-centre effects of its leader[12]. Specifically, the following gap decreases with the increase of lateral separation between the leader and follower. Xu et al present a car following model that takes into account the leading vehicle position and the centreline separation between two vehicles[13]. Tian et al analyse the car following behavioural stochasticity by measuring the wave travel time, and it is found that the follower can keep a nearly constant value of wave travel time, even if the leader's speed fluctuates significantly[14]. To sum up, these studies mainly analyse the influence of the leading vehicle on the longitudinal behaviours of the following vehicle, e.g. its longitudinal acceleration and velocity. However, in car following there are some interior couplings between longitudinal and lateral driving tasks, which has not been well studied yet.

On the other hand, for lateral driving automation, lane keeping or centring assist has been widely used in highway scenarios. There are basically three tasks to develop a lane keeping function, i.e., lane detecting, deciding a reference trajectory and tracking control. Most of lane keeping studies focus on the lane detection and trajectory tracking, while there are only few reports on how to choose a reference trajectory. Intuitively, the lane centreline is perhaps one best choice as the desired path. However, research shows that due to the shortage of accurate perception, drivers may have their own "lane centre", which is not necessarily the same to the real lane centreline [15]. Ding et al. also find that drivers' lane keeping behaviours may be influenced by several factors, e.g. lane position, disturbing objects and road curvature, and drivers tend to generate an "expected lateral driving zone" accordingly [16].

## 1.2. Personalization of ADAS functions

To achieve better customer experience and acceptance, the development team of ADAS must consider driver's preferences, which leads to the so-called personalized ADAS functions. In recent years, many studies focusing on driving personalization have been published, a part of which have been covered and discussed in an excellent review paper[17]. More recently, a design of a data-driven communication framework is proposed for personalized ADAS supports[18]. For specific ADAS functions, e.g. car-following or lane changing/keeping, there have been some personalization solutions. For example, a human-like car-following model is calibrated with driving simulator data, and a nonlinear model predictive controller is developed to reproduce the human-like behaviours in three different scenarios[19]. A lane change model is constructed by using modern statistical learning theory, which considers the static distribution and dynamic random process in driver behaviours[20]. In order to predict personalized driving behaviours, the driver-adaptive deep generative model and probabilistic approach are used to model the uncertainty of individual drivers in driving[21].

There are also a few publications on lane keeping personalization. For example, a learning-based model is presented to replicate the steering behaviour of expert drivers, in which 'motor intermittency' of human behaviour is introduced and the deep convolutional fuzzy system is used to build a feedforward-feedback control scheme[22]. Schnelle et al propose a combined driver model considering driver's desired path, which can fit the actual steering wheel angles in experiments [23]. Gao and Jiang use Artificial Potential Field to find the optimal preview trajectory for lane keeping, while the left and right lane marks can contribute differently to the weightings on potential forces [24]. An adaptive robust lane keeping controller is formulated on the basis of driver style dentification[25], and a rule based approach is used to classify driver style considering the levels of aggressiveness. In these approaches of lane keeping personalization, they incorporate driver preferences on the expected trajectory of ego vehicle, while the interaction of ego vehicle with the surrounding traffic is not considered.

As an intelligent combination of longitudinal and lateral driving automation, highway pilot (HWP) assist needs personalization, too. In essence, this should be accomplished with personalized interactions with road and surrounding traffic, which is true for both longitudinal and lateral sub-tasks. Existing researches on the highway pilot development mainly focus on driving safety, comfort or economy, while there are only limited reports on its personalization [26–28]. In addition, highway driving preferences should be considered more comprehensively, especially the influence of surrounding vehicle behaviours on the ego vehicle.

### 1.3. Main contributions

In this paper, a personalized highway pilot assist algorithm considering the leading vehicle behaviours is proposed. The main contributions of this study are three-fold.

(1) Based on both open dataset and simulator experiment studies, a driving phenomenon is reported, i.e. a non-negligible number of drivers in highway following driving choose their preferred reference trajectories under the influence of the leading vehicle behaviours.
(2) A preliminary speculation on the inherent mechanism is made, based on the results of both driving and eye-movement data.
(3) This understanding is further applied in the design of a personalized highway pilot assist algorithm, which includes the personalization of both speed control and lane keeping. Simulator experiments are conducted to validate our proposed algorithm, and results show significant reduction of subject driver mental workload.

The rest of this paper is organized as follows. Section 2 reports the phenomenon and presents the speculated mechanism, which works as both the motivation and theoretical foundation for the following personalization design. In Section 3, the algorithm design of personalized highway pilot assist is detailed, including the speed control, lane-keeping and personalization approaches. In Section 4, the proposed algorithm is validated in a driver-in-the-loop experiments, including driver mental workload and subjective evaluations. Finally, conclusions are given in Section 5.

## 2. Motivation: driving behaviour in car-following

### 2.1. Findings from naturalistic driving

The Highway Drone Dataset (highD) is a large-scale naturalistic vehicle trajectories dataset recorded at German highways [29]. It covers over 110 thousand vehicles, 44.5 driven kilometres and 147 driven hours. In our previous research on car following model, 664 following driving scenarios in one recording of highD dataset were analysed, with over 13,900 seconds total driven time of all tracked vehicles. If two vehicles keep driving in the same lane for more than ten seconds in this highway section, it is defined as car-following driving. Interestingly, approximately 40% of driver lane keeping behaviours in highway are affected by the lateral movement of the leading vehicle. Specifically, if leading vehicle ($V_l$) deviates to one lane marker, the following vehicle ($V_f$) will have the same trend of action. Fig. 1 shows a highway following driving case from highD, where the dashed red curve is the trajectory of $V_l$, while the solid blue curve is the trajectory of $V_f$. When $V_l$ starts moving to the left of the lane at $t_1$, $V_f$ moves to the same direction at $t_2$, after 0.96 second. And a same case appears at $t_3$ when $V_l$ starts moving in the opposite direction.

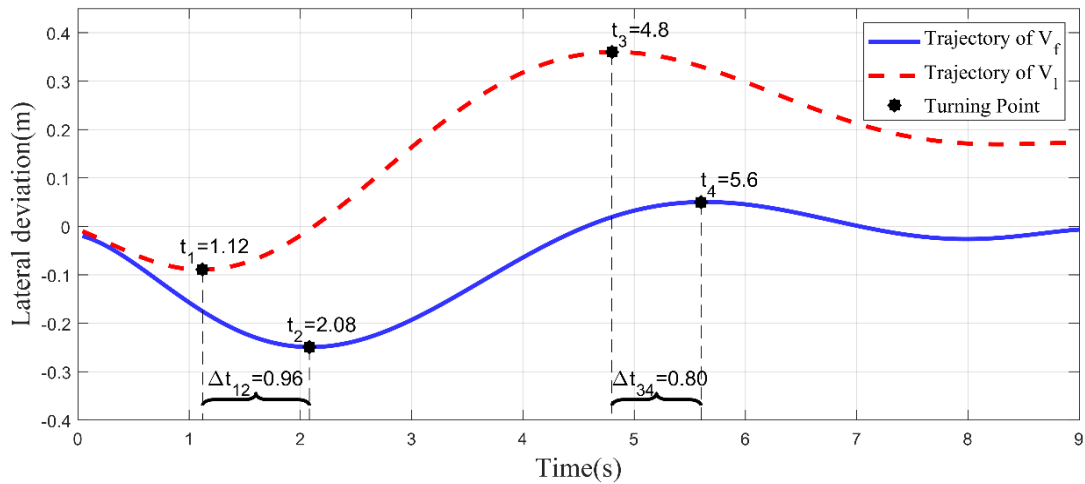

Fig. 1 A highway following driving case from highD dataset

Considering the large percentage of such phenomenon in the sampled highD data file, it is an interesting question whether this is a non-negligible phenomenon in highway driving. If yes, why do some drivers make such manoeuvre, while the others do not? How can this kind of behaviour be described with an interpretable model? Is it possible to integrate this personalization model into a highway pilot assist algorithm?

## 2.2. Findings from simulated driving

To answer the above questions, driver-in-the-loop experiments are conducted with a 6-Degree-of-Freedom (6-DoF) motion platform driving simulator, as shown in Fig. 2(a). The motion platform is controlled by a well-tuned motion cueing algorithm, and can provide a fidelity environment to the drivers during the experiment. IPG Carmaker software package is used to build the highway driving scenes, and the width of each lane is 3.75 meters, as shown in Fig. 2(b). Totally 16 domestic subject drivers are recruited to participate in this manual-driving experiment. They all have valid driving licenses and are aged between 20-30. Driving data is collected at default sample rate of 1ms, and drivers' eye movement data is collected by SMI ETG eye tracking glasses at 25Hz frequency.

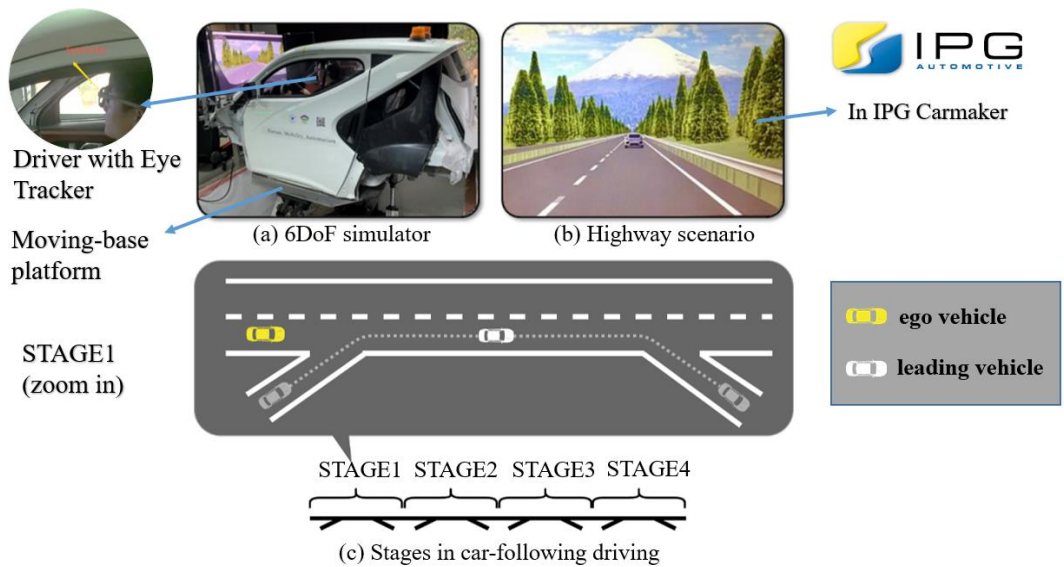

Fig. 2 Manual-driving experiment setup

This experiment includes two scenarios, free-driving and following-driving, while the traffic flow in the adjacent lane is set random with a determined density. In free-driving scenario, there is no leading vehicle ahead, and drivers are asked to keep driving in the right lane at a speed of less than 120 km/h. In the following-driving scenario, 4 leading vehicles with lateral behaviours are added. The following-driving scenario is divided into four stages with different leading vehicle speed: 90km/h, 100 km/h, 110 km/h, and 120 km/h, respectively. To make the scenario more realistic, the leading vehicle is arranged to ramps on and off in each stage, as shown in Fig. 2(c). In each following-driving stage, leading vehicle generates five lateral offsets with predefined displacements, i.e., 0.3m, 0.4m, 0.5m and 0.6m. The drivers need to drive in the right lane, and they are required to follow the leading vehicle without overtaking or lane change.

The experimental procedure is shown as follows.
1) Statement of experimental instructions;
2) Wear and calibrate the ETG eye tracking glasses;
3) Conduct experiment in free-driving scenario and collect data;
4) Sit relaxed for three minutes;
5) Re-calibrate the ETG eye tracking glasses, and conduct experiment in following-driving scenario;
6) Take off the eye tracking glasses and leave the simulator.

2.2.1. Driver gaze behaviours

Eye movement data are first used to analyse the subject drivers' gaze duration in three visual attention areas. Three areas of interest (AOIs) are focused, i.e., the instrument panel, the front and the lane markers, as shown in Fig. 3. It should be noted that in free-driving scenario, 'the front' means the far ahead area in from view, while in following-driving scenario, it means the area of leading vehicle. The statistical results of gaze duration for all 16 subject drivers are shown in Fig. 4. The white square/bar and the blue triangle/bar represent the statistical results in the free-driving and following-driving scenarios, respectively.

As shown in Fig. 4 (a), the gaze duration distribution of different drivers is dispersed in both scenarios. Fig. 4 (b) further shows that some drivers spend more time looking at the panel in both scenarios, e.g., subject drivers 6, 9 and 10, which leads to their lower proportions of gaze duration on lane-marker area, compared with subject drivers 0, 1 and 5. Statistically, in the car-following scenario, all 16 drivers pay more visual attention on the front leading vehicle than the lane marker, and 15 of them use the most visual attention on leading vehicle area. To sum up, the driver gaze duration data show that (1) drivers have their own personalized ways of allocating visual attention; (2) in car-following scenario, most drivers pay more visual attention on the leading vehicle than the other areas.

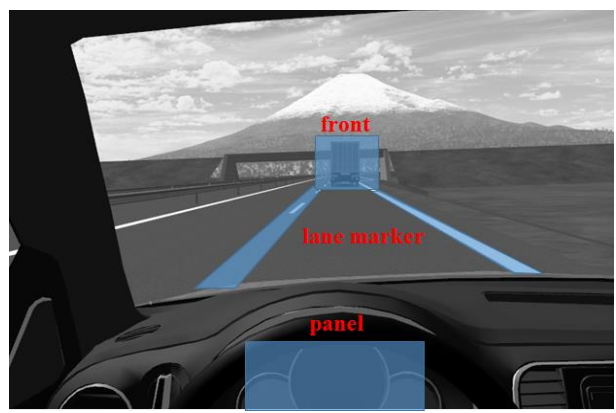

Fig. 3 Three areas of interest in the scene

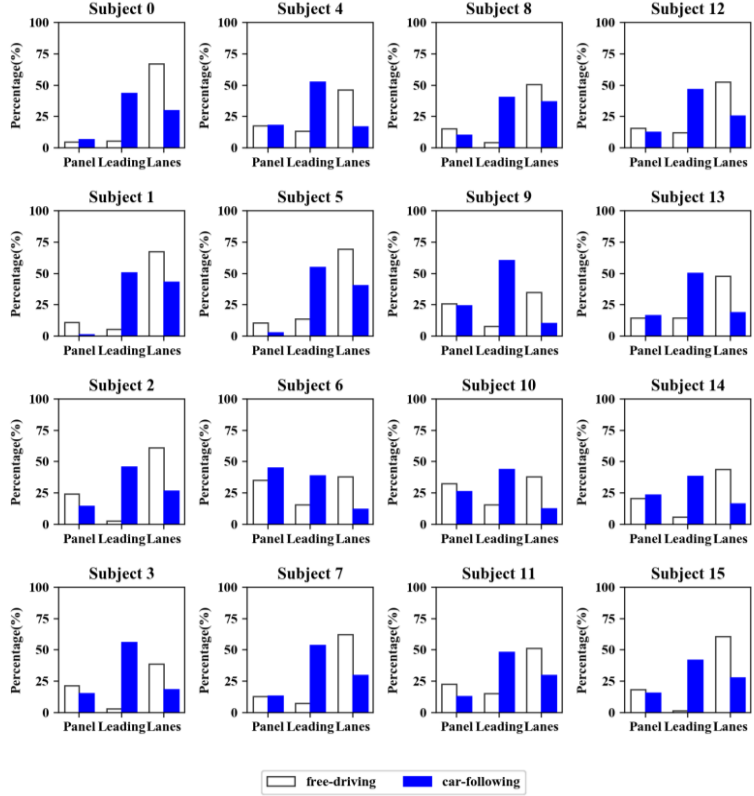

Fig. 4 Gaze duration distributions (a) and proportions (b) in different AOIs

2.2.2. Vehicle trajectory similarity

An intuitive way to judge whether a driver is affected during following driving, is to compare the ego vehicle's trajectory $Tr_e$ with the leading vehicle's trajectory $Tr_l$. If $Tr_e$ and $Tr_l$ have the similar shapes and trends, it means the driver is affected by the lateral behaviour of leading vehicle. Hausdorff Distance (HD) measures how close the shapes of two curves are, and it is widely used in comparing the similarity between different trajectories[30]. Here, HD between two trajectories $Tr_A$ and $Tr_B$ can be defined as follows.

$$H(Tr_A, Tr_B) = \max\bigl(h(Tr_A, Tr_B), h(Tr_B, Tr_A)\bigr) \quad (1)$$

where

$$h(Tr_A, Tr_B) = \max_{a \in Tr_A} \min_{b \in Tr_B} \|a - b\|. \quad (2)$$

The affected cases are selected if the distance $H(Tr_e, Tr_l)$ is greater than a reference distance, $H(Tr_{ref}, Tr_l)$. Here, the reference trajectory $Tr_{ref}$ is set as a straight line, while its lateral position is determined by the average lateral position of the points belonging to trajectory $Tr_l$.

There are 20 cases for one subject driver, each case corresponds to a specific speed and a front vehicle's lateral stimulus. To detail, four columns match four stages mentioned above and five rows match five lateral offsets generated by leading vehicle. Fig. 5 is an example of the judgment results about a subject driver. The solid red and dashed blue curves represent $Tr_e$ and $Tr_l$, respectively. For this driver, the affected cases are shown out with green background, i.e., case0, case7, case9, case14 and case19. The percentage of affected case for all drivers are shown in Fig. 6. After analysing all trajectories, it is found that for each subject driver there are at least two affected cases, which means that the percentage of affected cases is greater than 10%.

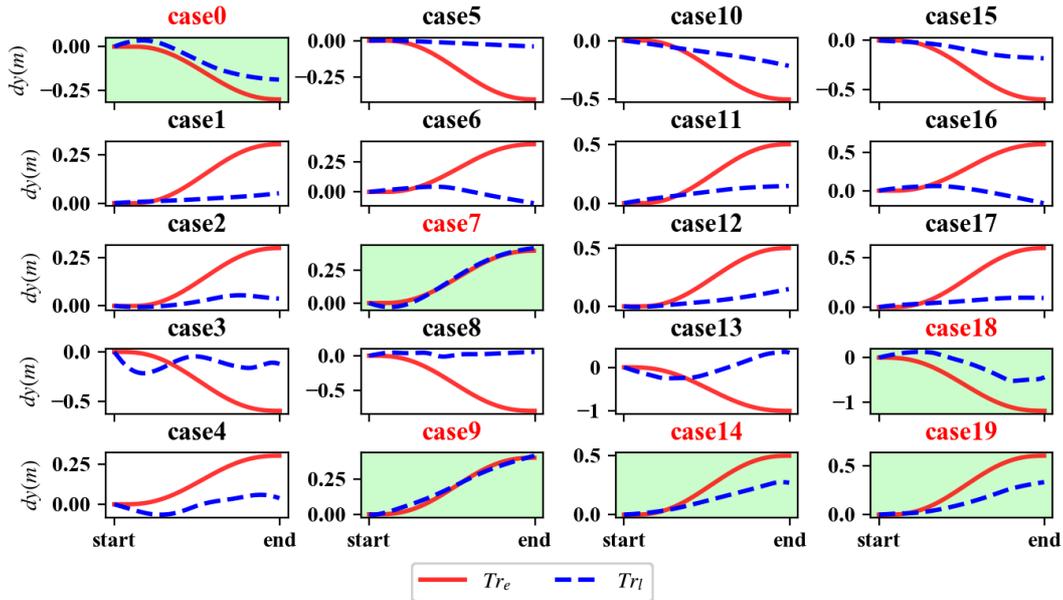

Fig. 5 Trajectory similarity judgement of one driver

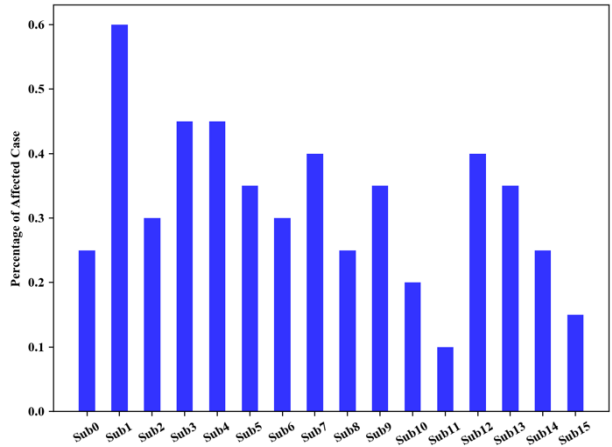

Fig. 6 Percentage of affected case for 16 drivers

## 2.3. Summary

The observations in both highD dataset and simulator experiments can lead to the following two findings.
(1) When following a leading vehicle, there is a non-negligible phenomenon that the lateral behaviours of following vehicle are affected by the leading vehicle.
(2) In following-driving scenario, drivers pay more attention to the leading vehicle, while if there is no leading vehicle in front, they spend more time observing lane markers. This may explain the phenomenon why drivers can be affected by the lateral behaviours of the leading vehicle. In the process of highway following driving, drivers handle lane-keeping manoeuvre mainly by referring to the leading vehicle, no matter whether they are aware of it or not.

In short, some drivers are more likely to be affected while others are not, and drivers also have different sensitivity and reaction time to the leading vehicle's lateral displacement. Therefore, these differences can represent the various driver preferences in highway following driving, which should be considered in ADAS personalization.

## 3. Algorithm design

On the basis of the previous findings, a personalized highway pilot assist algorithm is designed, with its structure shown in Fig. 7. For the longitudinal driving, a speed control model based on IDM is designed. For the lateral direction, a stimulus-response model is proposed to describe the driver's lane-keeping behaviour when following a leading vehicle. To be more specific, the leading vehicle lateral movement serves as the stimulus, while the ego vehicle's lateral decision serves as the response. Through the manual driving data collected in Section 2, three personalization parameters are extracted for every subject driver to personalize our algorithm.

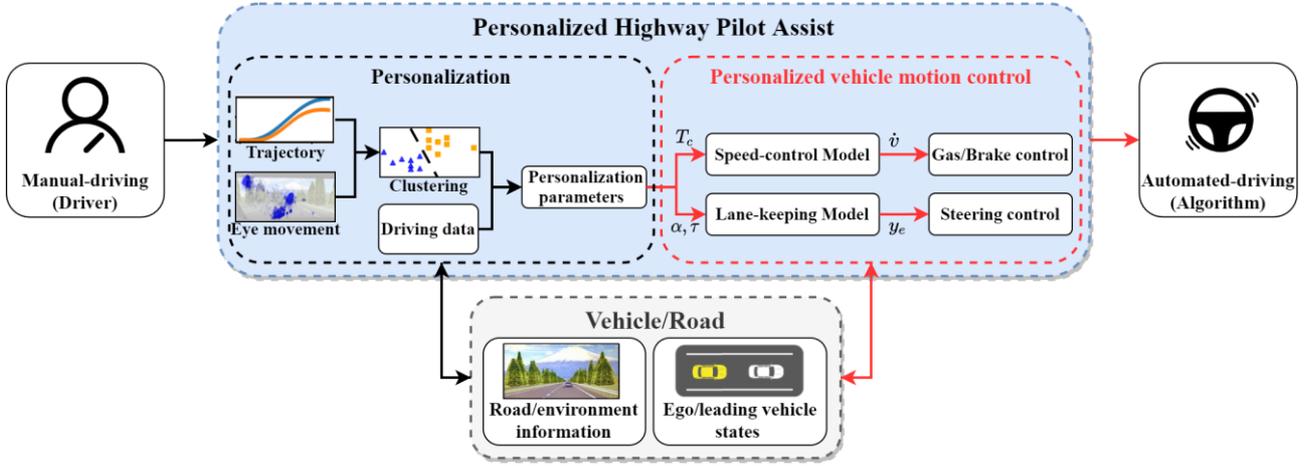

Fig. 7 The structure of the proposed highway pilot assist algorithm

### 3.1. Speed control with IDM

The IDM has the properties of collision avoidance and controllable stability [31], and it is a suitable model for the studied highway following driving scenario without any emergencies. The original IDM function is given by

$$\dot{v} = a\left[1 - \left(\frac{v}{v_0}\right)^\delta - \left(\frac{s^*(v, \Delta v)}{s}\right)^2\right] \qquad (3)$$

and

$$s_*(v, \Delta v) = s_0 + s_1\sqrt{\frac{v}{v_0}} + Tv + \frac{v\Delta v}{2\sqrt{ab}}. \qquad (4)$$

where there are seven physically meaningful parameters, i.e. desired velocity $v_0$, time headway $T$, maximum acceleration $a$, desired deceleration $b$, acceleration exponent δ, and jam distances $s_0$ and $s_1$. The inputs of IDM are the ego vehicle's velocity $v$, the gap $s$, and the velocity difference $\Delta v$ to the leading vehicle.

For speed control personalization, all the seven model parameters should be adapted to driver preferences and driving conditions. However, in this study, most of the following driving process are smooth, i.e. the leading vehicle travels at a constant desired speed without sudden acceleration or deceleration. Time headway $T$ is the most sensitive parameter to the drivers in our experiment. Therefore, here only a personalization time headway $T_p$ is fitted for each subject driver, while the remaining parameters in Table 1 are from the literature [7].

Table 1 IDM Parameters for speed control

| Parameter | Value |
| --- | --- |
| Desired velocity $v_0$ | 120 km/h |
| Time headway $T$ | $T_p$ |
| Maximum acceleration $a$ | 0.73 m/s² |
| Desired deceleration $b$ | 1.67 m/s² |
| Acceleration exponent $\delta$ | 4 |
| Jam distance $s_0$ | 2 m |
| Jam distance $s_1$ | 0 m |

### 3.2. Lane-keeping considering leading vehicle

As shown in Fig. 7, the lane-keeping model, as the core part, gives a desired lateral position $y_e$, which is to be tracked using steering control. The findings in naturalistic and simulated driving show that the following vehicle may move laterally in a similar way to the leading vehicle, but there may be differences of response time and magnitude sensitivity among different drivers. Therefore, in the design process of personalization, the driver sensitivity and response time are introduced. Then, driver's desired lateral position in ego vehicle at $k$ step, $y_e(k)$, can be derived from the previous lateral position $y_e(k-1)$ and the lateral displacement of leading vehicle at $k-d$ step, i.e. $\Delta y_l(k-d)$.

$$y_e(k) = y_e(k-1) + \alpha \Delta y_l(k-d) \tag{5}$$

and

$$d = \frac{\tau}{T_s}. \tag{6}$$

where $\alpha \in [0,1]$ and $\tau \in (0,2]$ are two continuously personalizable parameters. $\alpha$ describes the driver's sensitivity to the lateral stimulus of leading vehicle. If the driver does not care about the leading vehicle's lateral movement at all, it means $\alpha = 0$, while if the driver reproduces the leading vehicle lateral behaviour perfectly, then $\alpha = 1$. The reaction time $\tau$ stands for the response time delay after the driver observes the leading vehicle's lateral behaviour. And the constant $T_s$ is the sampling time during the driver-in-the-loop experiment. In this study, the calculation frequency is set as 50 Hz, hence $T_s = 0.02$.

To ensure safety, the ego vehicle is not allowed to cross the pre-set safe boundaries, as shown in the schematic diagram and lane-keeping flow chart in Fig. 8. In the example scenario, the width of ego vehicle is 2.1 meters, and the maximum lateral deviation of the leading vehicle is 0.6 meters. If the ego vehicle completely reproduces the leading vehicle's lateral behaviour ($\alpha$=1), its lateral coverage may reach 3.3 meters. To accommodate this, here the safety boundary width is set as 3.35m, i.e. the safe boundaries (dotted lines) are set 0.2m inside the lane boundaries (solid line). If there is no leading vehicle in front, the ego vehicle will keep driving on the centre-line (white dashed line). If there is a leading vehicle, and the ego vehicle is within safe boundaries, it will follow the desired lateral position given by Eq. (5), $y_e(k)$.

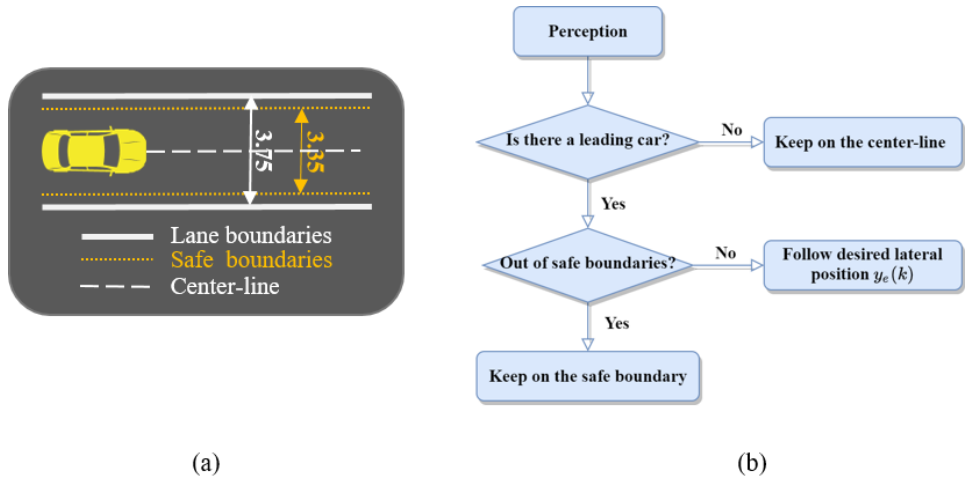

(a)                         (b)

Fig. 8 Schematic diagram of safe boundaries (a) and simplified lane-keeping logic (b)

### 3.3. Personalization

#### 3.3.1. Driver preference classification

Firstly, the driver preference classification is carried out, which later will work as the basis for determining the personalization parameters of our highway pilot assist algorithm. The data obtained in the simulator experiments are further used in classification, including the eye-movement data of subject drivers and the trajectories of leading/following vehicles.

The subject drivers are divided into two categories using the K-means clustering, which is an unsupervised machine learning method that aims to partition observation into $k$ clusters by minimizing the distance from the data point to the mean or median location of its assigned cluster [32]. The percentage of affected cases, $pc_a$, and the percentage of gaze times in leading area, $pc_g$, are considered as two feature dimensions in clustering. Fig. 9 summarizes the result of clustering, where the triangles in upper-right area correspond to the drivers with a driving style affected by the leading vehicle, denoted as $C_a$. The squares in lower-left area are the drivers with unaffected style, denoted as $C_u$. The red stars are the centres of two clusters, $C_a$ and $C_u$, respectively. The result shows that half drivers have the driving style that is affected by the leading vehicle's lateral movement.

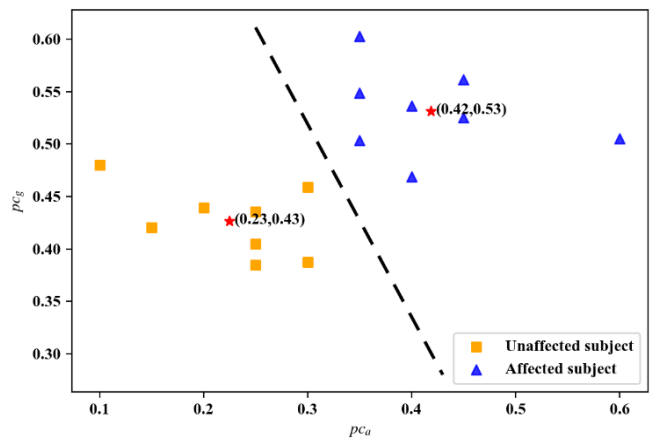

Fig. 9 Driving style clustering result of 16 drivers

Based on the driver style clustering results, analysis of variance (ANOVA) is adopted to test the significances of the impact of different AOIs between affected group $C_a$ and unaffected group $C_u$. The distributions of three AOIs are presented in Fig. 10, showing that in following-driving scenario, drivers in $C_a$ have a lower proportion of gaze duration in panel area ($M = 13.054$) than drivers in $C_u$ ($M = 19.302$) do, though not significant ($p = 0.24 > 0.01$). On the other side, the proportion of gaze duration in the leading vehicle area of drivers in $C_a$ ($M = 53.161$) is significantly higher than that of drivers in $C_u$ ($M = 42.641$), with $F = 31.554$, $p = 0.000063 < 0.01$. And as for the proportion of lanes area, no obvious difference is observed between the two groups. Therefore, to judge how likely the driver is affected by the leading vehicle's lateral behaviours, the driver visual perception characteristics, i.e. how much the driver gazes on the leading vehicle, can work as a good indicator. In practical applications, it can be used as a preliminary method to distinguish the two driver styles of the following vehicle.

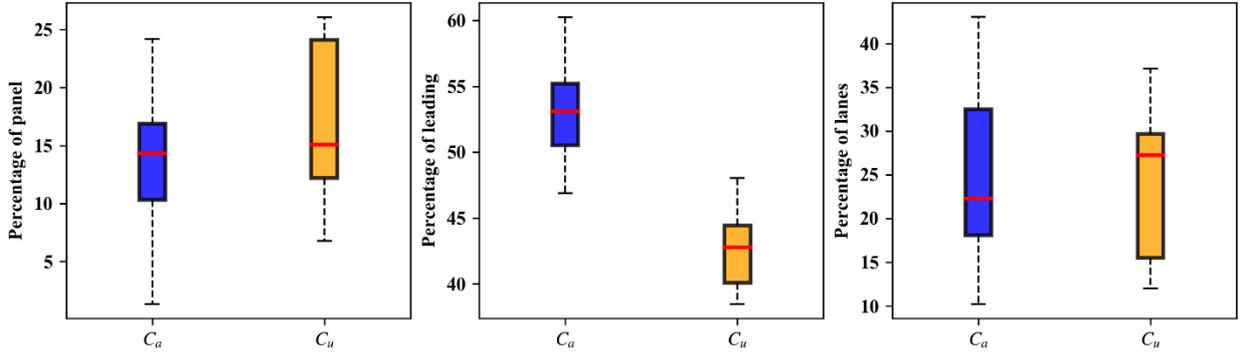

Fig. 10 Eye gaze statistics of different AOIs in following-driving scenario

3.3.2. Personalization parameters extraction

The personalization parameters of each subject driver need to be extracted to facilitate the application of the proposed algorithm. As for personalized speed control, time headways $T_p$ at different speeds are the average time headways of corresponding experiment stages. To personalize driver's lateral behaviour, the sensitivity parameter $\alpha$ and response time $\tau$ are discretised into a series of constant values with a step of 0.05. For each driver in affected group $C_a$, the ego vehicle's trajectories in affected cases (as shown in Fig. 5) are collected as the references. In each case, the optimal combination of $\alpha$ and $\tau$ is obtained by minimizing the squared error between the reference and the estimate trajectory generated by Eq.(5). The final personalization parameters for this driver are calculated by averaging the optimal combinations of the affected cases. While for the drivers in unaffected group $C_u$, they prefer to drive along the centreline of the lane without being affected, hence $\tau = 0$ and $\alpha = 0$. Finally, the personalization parameters for each subject driver are listed in Table 2.

Table 2 Personalization Parameters for Drivers

| Subject driver | $T_p$(90/100/110/120kph) (s) | $\tau$(s) | $\alpha$ | Affected ($C_a/C_u$) |
|---|---|---|---|---|
| 0 | 2.23/2.80/3.50/3.46 | 0. | 0. | $C_u$ |
| 1 | 1.50/1.11/0.85/1.25 | 1.030 | 0.913 | $C_a$ |
| 2 | 2.85/3.12/3.80/4.39 | 0. | 0. | $C_u$ |
| 3 | 2.05/2.72/3.25/3.75 | 0.628 | 0.367 | $C_a$ |
| 4 | 1.76/2.23/2.27/3.24 | 0.878 | 0.506 | $C_a$ |
| 5 | 1.46/2.04/2.13/2.74 | 1.264 | 0.593 | $C_a$ |
| 6 | 1.02/1.56/2.43/2.97 | 0. | 0. | $C_u$ |
| 7 | 2.44/2.95/3.10/3.51 | 1.263 | 0.869 | $C_a$ |
| 8 | 2.07/2.24/2.44/2.66 | 0. | 0. | $C_u$ |
| 9 | 1.93/2.47/3.07/3.59 | 0.679 | 0.350 | $C_a$ |
| 10 | 2.08/2.17/2.28/2.56 | 0. | 0. | $C_u$ |
| 11 | 1.82/2.04/2.62/2.88 | 0. | 0. | $C_u$ |
| 12 | 2.02/2.33/2.63/2.88 | 0.950 | 0.556 | $C_a$ |
| 13 | 1.78/2.07/2.22/2.53 | 1.293 | 0.793 | $C_a$ |
| 14 | 1.60/2.05/2.35/2.85 | 0. | 0. | $C_u$ |
| 15 | 2.22/2.48/3.03/3.16 | 0. | 0. | $C_u$ |

## 4. Experimental validation

To validate our personalization highway pilot assist algorithm, an automated-driving experiment is conducted. The proposed algorithm and two comparisons are applied in the following driving scenario. Two physiological metrics are selected to measure the mental workload during experiments, and the driver acceptance of algorithms are obtained by the self-report questionnaire.

### 4.1. Automated-driving experiment setup

The 16 drivers are invited again to the experiment. BioNomadix wireless electrocardiogram (ECG) sensor is installed, and the physiological signal is collected by BIOPAC MP160 and recorded by AcqKnowledge software. The drivers are asked to sit in the cockpit of the driving simulator and to concentrate on the car-following driving process.

The proposed algorithm, denoted $P$, and two comparison algorithms, denoted $C_1$ and $C_2$, are tested in a random order generated by a computer program. Each subject driver will evaluate three types of algorithms, (1) the Typical Type, i.e. Lane Centring assist, in which the vehicle keeps driving along the lane centreline; (2) the Following Type, in which the vehicle moves in the same direction as the leading vehicle; (3) the Opposite Type, in which the vehicle moves in the opposite direction to the leading vehicle. If the driver belongs to affected group $C_a$, algorithm $P$ is set to the Following Type and algorithm $C_1$ is set to the Typical Type. If the driver belongs to unaffected group $C_u$, algorithm $P$ is set to the Typical Type and algorithm $C_1$ is set to the Following Type. In detail, for any subject driver $S_i$, if $S_i \in C_a$, then $C_1$'s parameters, $\tau_{C1}$, $\alpha_{C1}$ are set to 0, and the $\tau_{C2}$, $\alpha_{C2}$ in $C_2$ were set to $\tau$ and $-\alpha$, respectively. As for $S_i \in C_u$, they have $\tau_{C1} = 1$, $\alpha_{C1} = 1$, and $\tau_{C2} = 1$, $\alpha_{C2} = -1$. Here, $\alpha > 0$ means the vehicle will move in the same direction as the leading vehicle, while $\alpha < 0$ means the vehicle will move in the opposite direction.

After each test the drivers are required to complete a self-report questionnaire and sit relaxed for three minutes to avoid affecting the next test.

The experimental procedure is shown as follows.

1) Enter in the simulator and listen to the statement of experimental instructions;
2) Fill in the basic information of subject driver on the tablet;
3) Wear the wireless ECG sensor;
4) Sit relaxed for three minutes;
5) Conduct experiment in following-driving scenario and collect data;
6) Complete a self-report questionnaire on the tablet;
7) Repeat steps 4)-6) twice;
8) Take off the ECG sensor and leave the simulator.

## 4.2. Subject driver mental workload results

Physiological signals such as electroencephalogram (EEG), electromyogram (EMG) and ECG, are increasingly applied to detect the status of drivers in various driving tasks. Some studies focus on the use of physiological signals to access driver mental workload [33,34]. Among various signals, ECG has been characterized as reliable and accurate indicator. Heart rate (HR) and heart rate variability (HRV) derived from ECG signals are suggested as two of the best correlate of driver workload [35].

In this study, HR and HRV analysis are used to show the mental workload of drivers, and further to evaluate the performance of different algorithms. The R-R interval (RRI), which is the interval between two adjacent R-waves on the ECG, and the ratio of the total energy in the low-frequency (LF) band (0-0.08Hz) and in the high-frequency (HF) band (0.15-0.5Hz), denoted as LF/HF, are the two metrics of the driver mental workload.

Previous researches have given reliable conclusions that, LF/HF will increase with increasing mental workload, while RRI will decrease. The statistical results about LF/HF and average RRI during three tests in automated driving experiment are given, and all 16 drivers are presented in Fig. 11, where three sets of bars correspond to three tests with $P$, $C_2$ and $C_2$, respectively. The white bars are the average RRIs and the blue bars represent the values of LF/HF.

As shown in Fig. 11, in general, drivers have higher RRI and lower LF/HF ratio in the test of their personalized algorithm $P$ rather than un-personalized algorithms $C_1$ and $C_2$. To further verify that the proposed algorithm can reduce the mental workload during following-driving process, ANOVA is used to test the significance between different algorithms. The distributions of two metrics are shown in Fig. 12. The RRI in the personalized algorithm $P(M = 0.914)$ is significantly larger than $C_1(M = 0.906)$ and $C_2(M = 0.898)$, $F = 9.640$, $p = 0.000328 < 0.01$. The LF/HF ratio in $P(M = 0.415)$ is significantly lower than $C_1(M = 0.919)$ and $C_2(M = 1.157)$, $F = 14.189$, $p = 0.001793 < 0.01$.

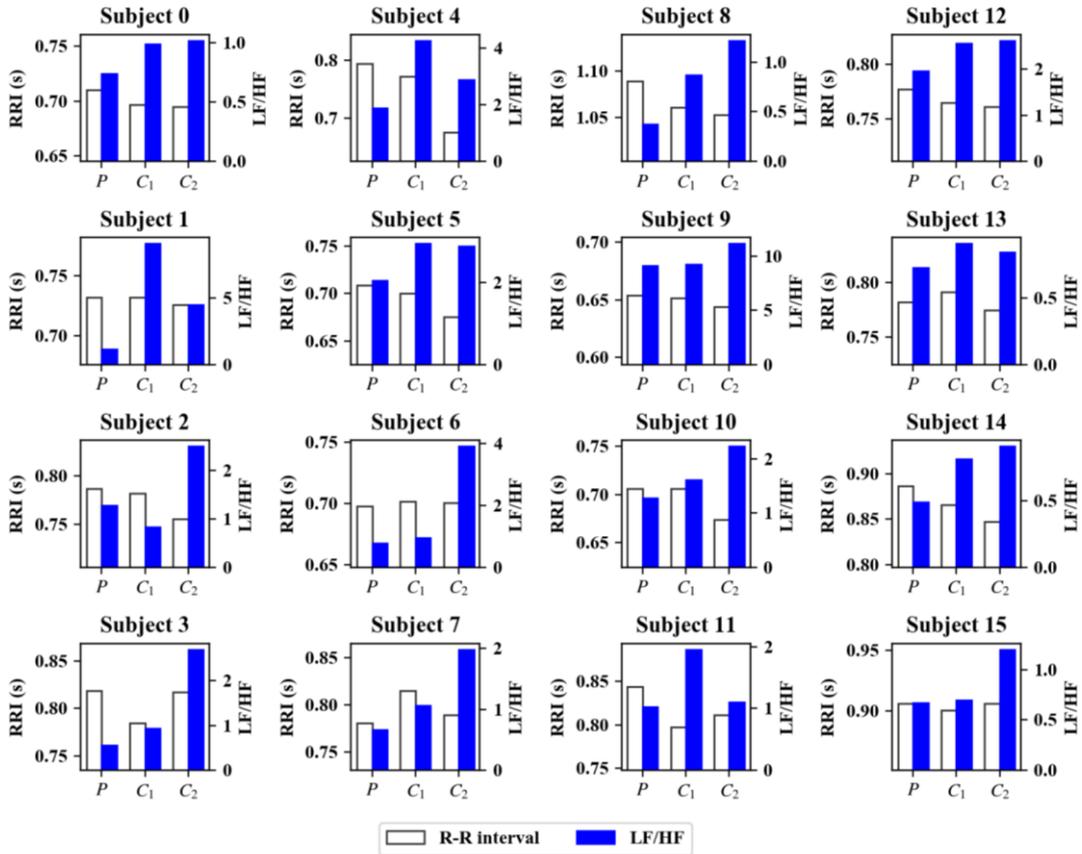

Fig. 11 Average RRI and LF/HF ratio of drivers

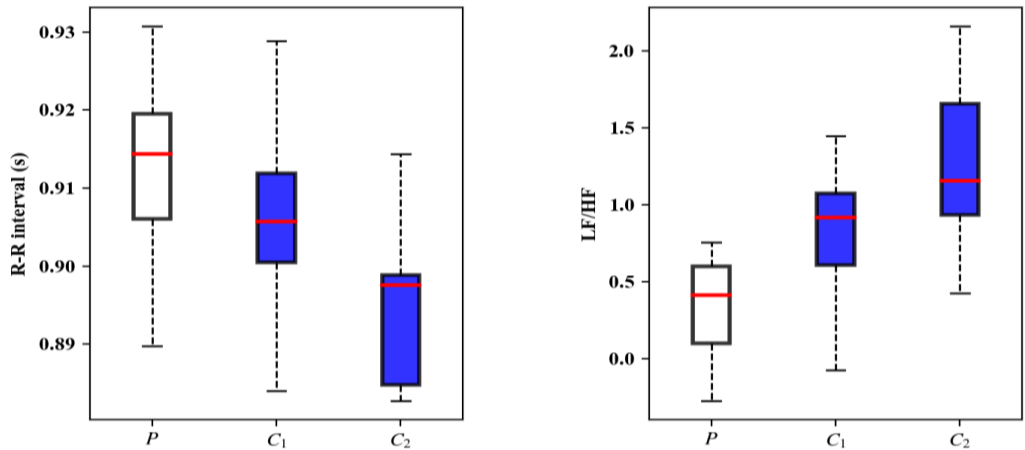

Fig. 12 ANOVA results of RRI and LF/HF ratio

### 4.3. Subjective evaluation

  Driver acceptance can be reflected in the statistical results of the self-report questionnaires. They are some multiple-choice questions in the questionnaire mainly involving the satisfaction survey of the algorithms, as detailed in Table 3. Drivers are asked to choose the option that best meet their feelings after every algorithm test. The statistics show that the average score of the personalized algorithm $P$ is 4.25, while that of the comparison algorithms, $C_1$ and $C_2$, are 3.33 and 3.08, respectively.

Table 3 User acceptance scale for algorithm evaluations

| Item | Grades | | | | |
|---|---|---|---|---|---|
| 1. Comprehension of the driving behaviours (bad~good) | 1 | 2 | 3 | 4 | 5 |
| 2. Acceptance of following distance (bad~good) | 1 | 2 | 3 | 4 | 5 |
| 3. Acceptance of lateral position (bad~good) | 1 | 2 | 3 | 4 | 5 |
| 4. Willingness to take over (strong~weak) | 1 | 2 | 3 | 4 | 5 |
| 5. Satisfaction with the automated driving algorithm (bad~good) | 1 | 2 | 3 | 4 | 5 |

**4.4. Discussion**

The results shown in Fig. 11 and Fig. 12 imply that, when the ego vehicle controlled by the proposed personalized highway pilot assist algorithm, most drivers feel more relaxed and have a lower mental workload during following driving process. This can be proved by their higher RRI and lower LF/HF ratio. The ANOVA results in Fig. 12 indicate that, compared with the $C_1$ and $C_2$, our proposed algorithm $P$ has a better performance with a significant reduction of mental workload ($p < 0.01$). In addition, the average score calculated from self-report questionnaire also shows that our proposed algorithm has a higher user acceptance. It is achieved by considering the driving style for every specific driver and presenting a personalized automation in highway following driving process. Hence, the advantage of our personalized highway pilot assist algorithm is validated through both objective physiological metrics and subjective evaluations.

More generally, it can be said that for the drivers in unaffected group $C_u$, i.e. those who prefer driving along the lane centreline or will not be affected by the leading vehicle, they will feel relaxed and satisfied if the vehicle always keeps driving in the centre of the lane. On the other hand, if the vehicle follows the lateral movements of the leading car, those drivers in affected group $C_a$ will benefit more. However, it should be noted that, for any subject driver, it is a bad decision to make the vehicle move in the opposite direction to the leading vehicle.

**5. Conclusion**

Starting from the unexpected findings when researching on the naturalistic trajectories in highD dataset, it is found that there are a non-negligible number of drivers whose lane keeping manoeuvre is obviously affected by leading vehicle's lateral behaviours. This phenomenon is observed again in simulator experiments, it may be because the drivers pay more attention on the leading vehicle, and their lane-keeping may mainly refer to the behaviour of leading vehicle. Inspired by the phenomenon and its underlying motivation, a personalized highway pilot assist algorithm is proposed to meet different drivers' driving styles. The driving styles are clustered by unsupervised machine learning and the personalization parameters for highway pilot assist algorithm are extracted after manual-driving experiments.

The proposed algorithm is validated by an automated driving experiment, showing that it can significantly reduce the subject driver mental workload ($p < 0.01$), compared with the un-personalized and oppositely-personalized algorithms. The self-reports from drivers also show a higher acceptance level of our proposed highway pilot assist algorithm.

In this paper, a simple highway following driving scenario is used for research, while in future study, our algorithm can be extended to more extensive scenarios, e.g., curve road, overtaking and on/off-ramp. For improvement, the degree of leading vehicle behaviour's influence on the following vehicle can be modelled probabilistically and integrated in the driving decision via Bayesian approaches. Additionally, this study only considers the leading vehicle's lateral behaviour in normal driving operations, while its lateral behaviours may also be caused by abnormal operations, e.g. driver's fatigue or drunk driving. In practical applications, such abnormal leading vehicle behaviours should be further handled in the algorithm strategies. The underlying mechanism of the observed driver behaviour can also be further explored, e.g. by combining brain science and psychology of decision.

## Nomenclatures

| Notation | Variable | Notation | Variable |
|---|---|---|---|
| $a$ | Maximum acceleration of IDM | IDM | Intelligent Driver Model |
| $b$ | Desired deceleration of IDM | LF | Total energy in the low-frequency |
| $pc_a$ | Percentage of affected cases | LF/HF | Ratio of LF and HF |
| $pc_g$ | Percentage of gaze times in front area | $P$ | Proposed algorithm |
| $s_0$ | Linear jam distance of IDM | RRI | R-R interval |
| $s_1$ | Non-linear jam distance of IDM | $S_i$ | Subject driver $i$ |
| $v_0$ | Desired velocity of IDM | $T$ | Time headway of IDM |
| $\dot{v}$ | Acceleration given by IDM | $T_p$ | Personalization time headway of IDM |
| $y_e$ | Desired lateral position of ego vehicle | $Tr_e$ | Ego vehicle's trajectory |
| $\Delta y_l$ | Lateral displacement of leading vehicle | $Tr_l$ | Leading vehicle's trajectory |
| AOI | Area of interest | $Tr_{ref}$ | Reference trajectory |
| $C_1$ | Comparison algorithm 1 | $T_s$ | Sampling time of experiment |
| $C_2$ | Comparison algorithm 2 | $V_f$ | Following vehicle |
| $C_a$ | Clusters of affected drivers | $V_l$ | Leading vehicle |
| $C_u$ | Clusters of unaffected drivers | $\alpha$ | Driver's sensitivity to the lateral stimulus of leading vehicle |
| ECG | Electrocardiogram | $\alpha_{c1}$ | Driver's sensitivity of $C_1$ |
| EEG | Electroencephalogram | $\alpha_{c2}$ | Driver's sensitivity of $C_2$ |
| EMG | Electromyogram | $\delta$ | Acceleration exponent of IDM |
| $H$ | Hausdorff Distance between two trajectories | $\tau$ | Driver's response time delay |
| HF | Total energy in the high-frequency | $\tau_{c1}$ | Driver's response time delay of $C_1$ |
| HR | Heart rate | $\tau_{c2}$ | Driver's response time delay of $C_2$ |
| HRV | Heart rate variability | | |

## Declaration of conflicting interests

The author(s) declared no potential conflicts of interest with respect to the research, authorship, and/or publication of this article.

## Funding

The author(s) disclosed receipt of the following financial support for the research, authorship, and/or publication of this article: This work was supported by Department of Science and Technology of Zhejiang (No. 2021C01SA601840, 2018C01058)

## ORCID iDs

Daofei Li, https://orcid.org/0000-0002-6909-0169
Ao Liu, https://orcid.org/0000-0001-5472-3253